\pdfoutput=1
\documentclass[sigconf]{acmart}

\usepackage[english]{babel}
\usepackage[utf8x]{inputenc}
\usepackage[T1]{fontenc}
\usepackage{amsthm}
\usepackage{paralist}
\usepackage{algorithm2e}

 \AtBeginDocument{%
  \providecommand\BibTeX{{%
    \normalfont B\kern-0.5em{\scshape i\kern-0.25em b}\kern-0.8em\TeX}}}

\usepackage{amsmath}
\usepackage{graphicx}
\usepackage[colorinlistoftodos]{todonotes}
\usepackage{hyperref}
\usepackage{subfig}
\usepackage{listings}
\usepackage{xparse}
\usepackage[super]{nth}

\theoremstyle{definition}

\theoremstyle{definition}

\theoremstyle{definition}

\theoremstyle{definition}

\theoremstyle{definition}

\setcopyright{none}
\renewcommand\footnotetextcopyrightpermission[1]{} 
\title{SFTM: Fast~Comparison of Web~Documents using Similarity-based~Flexible~Tree~Matching}
\author{Sacha BRISSET}
\affiliation{%
    \institution{Mantu}
    \country{France}
}
\email{sbrisset@mantu.com}

\author{Romain ROUVOY}
\orcid{0000-0003-1771-8791}
\affiliation{%
  \institution{Univ.\,Lille / Inria / IUF}
  \country{France}
}
\email{romain.rouvoy@univ-lille.fr}

\author{Renaud PAWLAK}
\affiliation{%
  \institution{Mantu}
    \country{France}
}
\email{rpawlak@mantu.com}

\author{Lionel SEINTURIER}
\orcid{0000-0003-1771-8791}
\affiliation{%
  \institution{Univ.\,Lille / Inria}
  \country{France}
}
\email{lionel.seinturier@univ-lille.fr}

\begin{abstract}
Tree matching techniques have been investigated in many fields, including web data mining and extraction, as a key component to analyze the content of web documents, existing tree matching approaches, like \emph{Tree-Edit Distance} (TED) or \emph{Flexible Tree Matching} (FTM), fail to scale beyond a few hundreds of nodes, which is far below the average complexity of existing web online documents and applications.

In this paper, we therefore propose a novel \emph{Similarity-based Flexible Tree Matching algorithm} (SFTM), which is the first algorithm to enable tree matching on real-life web documents with practical computation times.
In particular, we approach tree matching as an optimisation problem and we leverage node labels and local topology similarity in order to avoid any combinatorial explosion.
Our practical evaluation demonstrates that our approach compares to the reference implementation of TED qualitatively, while improving the computation times by two orders of magnitude.
\end{abstract}

\begin{document}
\maketitle

\newcommand{\FTM}{\mathrm{FTM}}
\newcommand{\SFTM}{\mathrm{SFTM}}
\settopmatter{printacmref=false}
\renewcommand\footnotetextcopyrightpermission[1]{} 
\pagestyle{plain} %


\section{Introduction} 
The success of Internet has lead to the publication and the delivery of a deluge of web documents.
In particular, web services and applications are heavily using XML and JSON standards to transfer information across the network as structured web documents.
Inevitably, the success of these technologies has led to the definition of more complex and large web documents that keep evolving over time.
However, keeping track of such changes remains a critical issue for the ecosystem and the research community.
Examples of usages that require to detect or track changes in web documents include Web extraction~\cite{reis2004automatic,yao2013answer,zhai2005web}, Web testing~\cite{stocco2017apogen, choudhary2011water}, comparison of Web service versions~\cite{fokaefs2011empirical}, Web schema matching~\cite{hao2007web} and automatic re-organization of websites~\cite{Kumar2011_Bricolage}.

\begin{figure}
	\includegraphics[width=.8\linewidth]{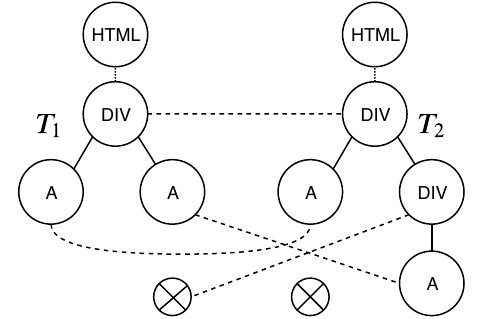}
    \caption{Example of a tree matching. Crossed circles are auxiliary \emph{no-match} nodes enabling insertions and removals between trees.}
    \label{fig:tree_matching_example}
\end{figure}

As all of these usages share the need to compare \emph{Document Object Model} (DOM) trees across multiples versions of a web document, several algorithms have been proposed to achieve this goal.
The traditional approach, which is general to any kind of tree, is \emph{Tree Edit Distance} (TED)~\cite{Tai1979} and can be computed in minimum $O(N^3)$ time~\cite{bringmann2018tree}. Figure \ref{fig:tree_matching_example} illustrates the tree-matching problem. 
TED is a restriction of tree matching where descendants of matched nodes can only match with each others (ancestry restriction) and siblings order must be preserved.
We executed a robust implementation of TED, named APTED~\cite{pawlik2015efficient,pawlik2016tree}, on two instances of the DOM of YouTube, which took more than 4 minutes to propose a matching.
Unfortunately, when processing and comparing a large dataset of Web documents, one cannot afford  such computation times, which makes TED difficult to use in production.

The qualitative restrictions and speed limitations of TED therefore led to the development of alternative algorithms.
\cite{fokaefs2011empirical} extended TED with some additional move operations executed \emph{a~posteriori} to address the ancestry restriction.
\cite{Kumar2011_FTM,Kumar2011_Bricolage} developed her own \emph{Flexible Tree Matching} (FTM) algorithm to address the ancestry restriction problem, while \cite{reis2004automatic} developed a fast matching system based on top-down matching to extract news faster than TED does. 

In the line of the aforementioned work, this paper aims at enabling the fast and non-restricted comparison of complex web documents.
We propose an extended version of FTM, named \emph{Similarity-based Flexible Tree Matching} (SFTM), that leverages similarity metrics to speed up the comparison.
SFTM retains the advantage of FTM to offer a non-restricted tree matching while offering computation times much lower than even restricted versions of the problem.
The algorithm exposes performance parameters to trade computation time and matching accuracy. To the best of our knowledge, SFTM is the first solution to match real-life web documents in practical time (e.g. SFTM matches the DOM of Youtube in less than a second compared to 4 minutes for APTED).
Through empirical evaluation on real websites, we show that---for selected parameters---our implementation of SFTM qualitatively compares to APTED and empirically seems to scale in $O(n\cdot log(n))$ with the size of the considered DOM, thus making it applicable in many production contexts.

The remainder of this paper is organized as follows.
Section~\ref{sec:related_work} covers the related work.
Section~\ref{sec:ftm} introduces the \emph{Flexible Tree Matching} (FTM) original algorithm.
Section~\ref{sec:SFTM} presents \emph{Similarity-based Flexible Tree Matching} (SFTM), our extension of FTM that leverages the node labels and local topology similarity to guide the comparison.
Section~\ref{sec:evaluation} thoroughly evaluates our solution against the state of the art on a realistic dataset of web documents.
Section~\ref{sec:threats} discusses the threats to validity of our contribution.
Section~\ref{sec:conclusion} concludes and overviews some perspectives for this work.

\section{Related Work}\label{sec:related_work}
\paragraph{Tree Edit Distance (TED)}\label{sec:ted}
Comparing two trees is a problem that has been at the center of a significant amount of research.
In 1979, Tai~\cite{Tai1979} introduced the \emph{Tree Edit Distance} (TED) as a generalization of the standard \emph{edit distance} problem applied to strings.
Given two ordered labeled trees $T_1$ and $T_2$, the TED is defined as the minimal amount of node insertion, removal or relabel to transform $T_1$ into $T_2$, while different cost coefficients can be assigned to each type of operation.
By following an optimal sequence of operations applied to $T_1$, it is possible to match the nodes between $T_1$ and $T_2$.
This problem has been extensively studied since then to reduce the space and time complexity of the algorithm that computes the TED.
To the best of our knowledge, the reference implementation available today is the \emph{All-Path Tree Edit Distance} (APTED)~\cite{Pawlik2011, pawlik2015efficient, pawlik2016tree} with a complexity of $O(n^2)$ in space and $O(n^3)$ in time in the worst case, where $n$ is the total number of nodes ($n = |T_1|+|T_2|$).
In our work, we consider APTED as the baseline to evaluate our contribution.

\cite{bringmann2018tree} showed that TED cannot be computed in worst case complexity lower than $O(n^3)$.
In order to circumvent this limitation, several restricted versions of the TED problem have been formulated.
The \textit{Constrained Edit Distance}~\cite{zhang1995algorithms, zhang1996constrained} is an edit distance where disjoint subtrees can only be mapped to disjoint subtrees.
The \textit{Tree Alignment Distance}~\cite{jiang1994alignment} is a TED where all insertions must be performed before any deletion.
The \textit{Top-Down} distance~\cite{selkow1977tree} is computable in $O(|T_1||T_2|)$, but imposes as a restriction that the parents of nodes in a mapping must be in the mapping.
The \textit{Bottom-Up} distance~\cite{valiente2001efficient} between trees allows to build a mapping in linear time, but such mapping must respect the following constraint: if two nodes have been mapped, their respective children must also be part of the mapping.
\cite{reis2004automatic} proposes a variation of the \textit{Top-Down} mapping, called \textit{Restricted Top-Down Mapping} (RTDM), where replacement operations are restricted to the leaves of the trees, which delivers considerable speed gains, despite a theoretical worst case time complexity still in $O(N^2)$. By definition TED already sets strong restrictions on produced matchings: sibling order and ancestry relationships must be preserved \cite{zhang1995algorithms}. These restrictions are particularly problematic when matching two full web documents together~\cite{Kumar2011_Bricolage}. While above solutions improve computation times, they answer a restricted version of the TED problem leading to an even more restricted set of possible matchings.

\paragraph{Flexible Tree Matching (FTM)}
In~\cite{Kumar2011_Bricolage}, TED is found to be unpractical when applied on DOM, as the resulting matching enforces ancestry relationship---\emph{i.e.}, once $n \in T_1$ and $m \in T_2$ have been matched, the descendants of $n$ can only be matched with the descendants of $m$, and \emph{vice~versa}. 
Consequently, Kumar~\emph{et~al.} introduced the notion of \textit{Flexible Tree Matching} (FTM), which relaxes the ancestry relationship constraint at the price of a strong complexity.
It restricts its use to small HTML trees composed of hundreds of nodes, thus making it unpractical for modern web documents, often including thousands of nodes.

We therefore aim at reducing the complexity of the FTM algorithm in order to scale on complex web documents without enforcing restrictions on produced tree-matching solutions.
More specifically, our contributions read as follows:
\begin{compactenum}
    \item We develop an extended FTM algorithm, coined as \emph{Similarity-based Flexible Tree Matching} (SFTM), by leveraging the notion of label similarity, and similarity propagation to reduce the computation time,
    \item We apply mutations on real-life web documents, and provide a thorough evaluation of our implementation of SFTM, showing it outperforms state-of-the-art approaches in terms of scalability and performance, yet offering similar qualitative results.
\end{compactenum}

\section{Flexible Tree Matching}\label{sec:ftm}
The \emph{Similarity-based Flexible Tree Matching} (SFTM) we introduce in this paper is an extension of the \textit{Flexible Tree Matching Algorithm} (FTM).
This section therefore introduces the FTM algorithm, as originally proposed by Kumar~\emph{et~al.}~\cite{Kumar2011_Bricolage}.
We first describe the notations used throughout the rest of the paper, and then describe the main steps of the algorithm.

Building on the terminology from~\cite{Kumar2011_FTM}, we consider a matching between two labeled trees $T_1$ and $T_2$ comprising $|T_1|$ and $|T_2|$~nodes, respectively.
We note $N = max(|T_1|, |T_2|)$.

\begin{figure}
    \includegraphics[width=.8\linewidth]{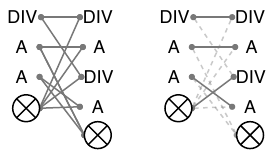}
    \caption{From the input trees depicted in Figure~\ref{fig:tree_matching_example}, we build a bipartite graph $G$ representing the set of all possible matching (left) and then compute the optimal full matching (right).}
    \label{fig:g_SFTM}%
\end{figure}

Let us consider the complete bipartite graph $G$ between $T_1^* = T_1\cup{\Theta _1}$ and $T_2^* = T_2\cup{\Theta _2}$, where $\Theta_1$ and $\Theta_2$ are \textit{no-match} nodes.
The fact that $G$ is complete means that every nodes of $T_1^*$ shares exactly one edge with every nodes of $T_2^*$.
An edge $e(n, m) \in E(G)$ between $n\in T_1^*$ and $m\in T_2^*$ represents the matching of $n$ with $m$.
So, intuitively, $G$ represents all possible matchings between $T_1^*$ and $T_2^*$ (cf. Figure~\ref{fig:g_SFTM}).
We call \textit{matching} and note $M\subset E(G)$, a subset of edges selected from $G$.
A matching $M$ is said to be \textit{full} \emph{iff} each node in $T_1$ has exactly one edge in $M$ that links it to a node in $T_2^*$ and, inversely, each node in $T_2$ has exactly one edge in $M$ that links it to a node in $T_1^*$.
Since matchings need to be \textit{full}, the auxiliary \textit{no-match} nodes $\Theta_1, \Theta_2$ are needed to allow insertion and deletion operations.
The set of possible \textit{full} matchings is restricted to the set of matchings $\mathcal{M}$ satisfying that every node in $T_1 \cup T_2$ is covered by exactly one edge.
\emph{No-match} nodes are the only nodes allowed to be involved in multiple edges.

Given an edge $e(n, m)\in E(G)$ linking $n$ to $m$, FTM defines the cost $c(e)$ to quantify how different $n$ and $m$ are, considering both their labels and the topology of the tree.
Starting from the bipartite graph $G$ describing all possible matchings, the idea behind FTM is to compute the costs $c(e)$ of each edge $e\in E(G)$ and to find the optimal matching with respect to costs---\emph{i.e.}, to select the set of edges $M\subset E(G)$, such that $M$ is \textit{full} and $c(M)$ is minimal (where $c(M) = \sum_{e\in E(G)}c(e)$).

\begin{figure*}
	\includegraphics[width=.8\linewidth]{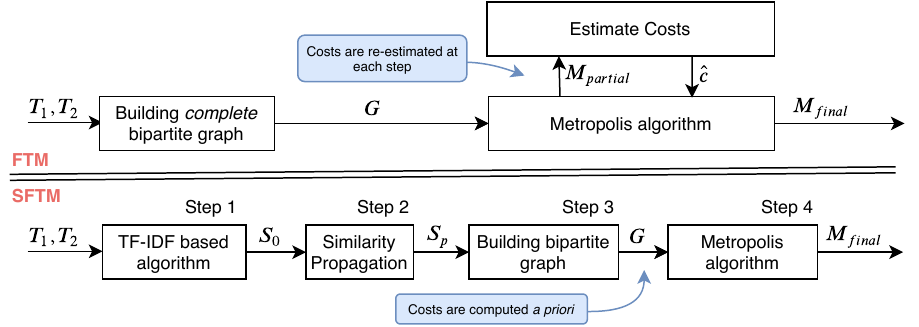}
    \caption{Steps to compute a full matching between two tree $T_1$ and $T_2$. In the top, we describe FTM and in the bottom, our algorithm: SFTM}
    \label{fig:steps}
\end{figure*}

The upper part of the Figure~\ref{fig:steps} describes the main steps involved in computing the final full matching between $T_1$ and $T_2$.

\subsection{Cost Estimation}\label{sec:FTM_cost}
As FTM provides a wide flexibility regarding possible matchings, the design of the cost function is a key parameter in order to obtain a matching that takes into account both the labels and the topology of the trees.
Typically, the cost of an edge $e$ between two nodes $n$ and $m$ is computed as follows:
\begin{equation}\label{eq:FTM_cost}
c(e) =
\begin{cases}
    w_n                                  &\text{if}\ n\ or\ m \in \{\Theta_1, \Theta_2\} \\
    w_r c_r(e) + w_a c_a(e) + w_s c_s(e) & \text{otherwise}
\end{cases}
\end{equation} 
where $\Theta_1, \Theta_2$ are \textit{no-match} nodes, $w_n$ is the penalty when failing to match one of the edge ends, $c_r$, $c_a$ and $c_s$ are the cost of \textit{relabeling}, \textit{violating ancestry relationship} and \textit{violating sibling group}, respectively, and $w_r$, $w_r$ and $w_r$ their associated weight in the cost function.
$w_n, w_r, c_r, w_a$ and $w_s$ are parameters of the cost function that depend on the kind of matching the user requires.
By extension, we note $c(M) = \sum_{e \in M} c(e)$ the cost of a matching $M$.

Given $e(n, m)$, the ancestry and sibling costs, $c_a(e)$ and $c_s(e)$, model the changes in topology that matching $n$ with $m$ entails.
Unfortunately, we can only compute the costs $c_a$ and $c_s$ if we have access to a full matching, as both costs require a knowledge on how other nodes in the tree were matched (\emph{e.g.}, $c_a$ involves counting the number of children of $n$ matched with nodes that are not children of $m$).
In order to circumvent the problem, FTM defines the approximate costs $\hat{c_a}, \hat{c_s}$ that can be computed from bounds on the different components of the cost $c$.
Practically, in order to generate one possible full matching, FTM iteratively selects edges in $G$ and, each time an edge is selected, the bounds of $c$ are tightened (we can approximate $c$ more precisely), which means the costs $\hat{c_a}, \hat{c_s}$ must be recomputed.
This is illustrated in the upper part of Figure~\ref{fig:steps}. 

The need to recompute the approximated costs after each edge selection therefore imposes some critical limitation on the scalability of the algorithm.

\subsection{Metropolis Algorithm}\label{sec:metropolis_ftm}%
Finding the optimal matching, given the graph $G$ and the cost function $c$ is a challenging problem, the authors even proved in~\cite{Kumar2011_FTM} that the problem is NP-hard.
Consequently, the authors described how to use the Metropolis algorithm~\cite{metropolis1953equation} to approximate the optimal matching. 
The Metropolis algorithm provides a way to explore a probability distribution by random walking through samples.
FTM uses this algorithm to random walk through several full matchings, and select the least costly.
The algorithm needs to be configured with:
\begin{compactenum}
    \item An initial sample (full matching) $M_0$,
    \item A suggestion function $M_t \mapsto M_{t+1}$,
    \item An objective function to maximize: $f: M \mapsto \text{quality of } M$,
    \item The number of random walks before returning the best value.
\end{compactenum}

Kumar~\emph{et~al.} defines the objective function $f$ by:
\begin{equation} \label{eq:objective_FTM}
	f_{FTM}(M) = \exp(-\beta\ c(M))
\end{equation}
In order to suggest a matching $M_{t+1}$ from a previously accepted one $M_t$, FTM selects a random number of edges from $M_t$ to keep, sorts remaining edges by increasing costs and iterate through the ordered edges with a chance $\gamma$ to select it.
Once an edge $e(n,m)$ is selected, all edges connected to $n$ and $m$ are removed from $G$, approximate costs need to be recomputed for all edges and sorted so we can select another edge.
The process is repeated until a full matching is obtained.

Despite using the Metropolis algorithm to reduce the time complexity of the problem, the overall algorithm remains prohibitively costly to compute (cf. Section~\ref{sec:evaluation}), notably due to the continuous computation of the approximated cost for each step of the full matching generation.

\subsection{Complexity Analysis}\label{sec:ftm_complexity}
The original FTM paper~\cite{Kumar2011_FTM} does not provide any information on the complexity or the computation time of the algorithm.
We provide an analysis of FTM's theoretical complexity to use as a baseline to our approach (SFTM).

\paragraph{Complete bipartite graph $G$}
Building the complete bipartite graph requires linking each node form $T_1$ to each node from $T_2$, which requires $O(N^2)$ operations. 

\paragraph{Metropolis Algorithm}
For each iteration of the Metropolis algorithm, FTM needs to suggest a new matching.
In the worst case, the algorithm should choose among all $N^2$ edges.
Each time an edge between $e_1$ and $e_2$ is selected, all other edges connected to $e_1$ and $e_2$ are pruned and costs a re-estimated.
It means that costs need to be re-computed and sorted for $N^2$ edges, then $(N-1)^2$ edges (after selection and pruning) and so on until all edges have been selected or pruned.
This implies that the total number of times the costs are re-computed and sorted is in $O(\sum^N_{n = 0}n^2)$ = $O(N^3)$.
Computing the cost for a given edge linking $e_1$ and $e_2$ involves counting the number of potential ancestry and sibling violations, which requires going through all edges connected to siblings and children of $e_1$ and $e_2$.
Even if we assume the number of siblings and children is independent of $N$, it still means estimating the cost of one edge requires $O(N)$ operations.
Thus, in the worst case, the amount of operations done by FTM for each iteration of the Metropolis algorithm is in $O(\sum^N_{n = 0}n^3)$ = $O(N^4)$ (using Faulhaber's Formula).

\section{Similarity-based Flexible Tree~Matching}\label{sec:SFTM}
\emph{Similarity-based Flexible Tree Matching} (SFTM) replaces the cost system of FTM by a similarity-based cost that can be computed \textit{a~priori}.
This approach drastically improves computation times and exposes a parameter that can be tuned to find the desired trade-off between computation time and matching accuracy.

Given two trees $T_1$ and $T_2$, SFTM relies on the creation of a \textit{similarity metric} between the nodes of $T_1$ and $T_2$.
We compute this similarity metric for all pairs of nodes using \emph{i)} inverted indices for labels and \emph{ii)} label propagation for the topology.
We build a bipartite graph $G$ using this similarity metric to compute the costs and apply the Metropolis algorithm to approximate the optimal full matching from $G$.
This new similarity measure allows us to improve the FTM algorithm in two key aspects:
\begin{compactenum}
	\item When building $G$, we do not create all $N^2$ possible edges. We only consider edges linking two nodes with a non-null similarity.
    \item When generating a full matching, we never need to recompute the costs since these costs are solely dependant on our similarity measure.
\end{compactenum}
In this section, we
\begin{inparaenum}[(a)]
	\item introduce our new similarity metric and
    \item describe how we leverage it to approximate the optimal full matching.
\end{inparaenum}

\subsection{Node Similarity}\label{se:newCost}
The similarity metric between nodes from $T_1$ and $T_2$ is computed in two steps:
\begin{inparaenum}
	\item we compute $S_0$, the initial similarity function using only \textit{labels} of the trees individually, and then
    \item we transform $S_0$ to take into account the topology of the tree and compute our final similarity function $S_p$.
\end{inparaenum}
The computation of $S_0$ leverages inverted index techniques traditionally used to query text in a large document databases.
In our case, documents we query against are nodes from $T_1$ and queries are extracted from $T_2$ nodes.

\subsubsection{Initial Similarity (step 1)}
To compute the initial similarity $S_0$ (\textit{step 1} in Figure~\ref{fig:steps}) between $T_1$ and $T_2$, we independently compare the labels of $T_1$ and $T_2$ using TF-IDF.
The resulting initial similarity $S_0$ does not take the topology of the trees into account.

In order to take into account relabeling cost between nodes, FTM and TED allow the user to input a pairwise comparison function $label(n), label(m) \mapsto similarity\ score$.
Computing this similarity score for all the pairs of nodes requires $O(N^2)$ operations.
To reduce the number of operations, SFTM uses---instead---inverted indices: we require the user to input a $tokenize : n \mapsto \ token\ list$ function, and then
\begin{inparaenum}
	\item we sort each node $n$ from $T_1$ into a set of tokens (as defined by the $tokenize$ function), before
    \item we iterate through tokens of nodes $m$ from $T_2$ and increase the value of $S_0(n,m)$ for each token $n$ and $m$ have in common.
\end{inparaenum}
Section \ref{tokenSelection} provides a detailed description of the $tokenize$ function we use in our evaluation.

After sorting nodes $n$ from $T_1$ into tokens, we obtain an inverted index $T_{map}$ (for Token Map), which is a table where each entry contains one token $t$ along with the list of nodes that contains the token.

The idea behind the inverted index $T_{map}$ is to use the information that a node $n \in T_1$ belongs to a token as a differentiating feature of $n$ allowing to compare it to nodes $m \in T_2$.
If a token contains all nodes in $T_1$, this token has no differentiating power.
In general, the rarest a token, the more differentiating it is.
This idea is very common in \emph{Natural Language Processing} (NLP) and a common tool to measure how rare is a token is TF-IDF and more precisely, the \emph{Inverted Document Frequency} (IDF) part of the formula.

Applying TF-IDF to our similarity yields the following definition:
\begin{align}
	IDF(t) &= log(|T_1|/|T_{map}[t]|) \\
	S_0(n,m) &= \sum_{t\in tk(n) \cap tk(m)}IDF(t)
\end{align}
The \emph{Inverted Document Frequency function} (IDF) is a measure of how rare a token is, $|T_{map}[t]|$ is the number of nodes containing the token $t$ and $tk$ is a short for the user input $tokenize$ function.
Intuitively, we retrieve all common tokens between $n$ and $m$, and for each common token $t$, we increase $S(n,m)$ by a high value if $t$ is rare and a low value if $t$ is common.
In Section \ref{sec:implementation}, we give a detailed implementation of how to compute the initial similarity $S_0$.

Tokens that appear in many nodes have little impact on the final score (\emph{i.e.}, low IDF) yet have a very negative impact on the computation time.
In our algorithm, we expose the sublinear threshold function $f: N \mapsto f(N) < N$ as a parameter of the algorithm.
We use $f$ to filter out all tokens appearing in more than $f(N)$ nodes.
$f$ defines a threshold between computation time and matching quality: when $N -f(N)$ decreases, computation times and matching quality increase.
In Section~\ref{sec:complexity}, we discuss how $f(N)$ influences the worst-case theoretical complexity.

\subsubsection{Local Topology (step 2)}
$S_{0}$ represents the similarity between node labels, but does not take into account the topology of the trees.
To weight in local topology similarities, we propagate the score of each node couple to their offspring.
This idea of propagation is inspired by recent \emph{Graph Convolutional Network} (GCN) techniques~\cite{kipf2016semi}.

The original FTM algorithm includes two terms in the cost function, $c_a$ and $c_s$, which reflect the topology of the trees.
Since we do not use these terms, we need our similarity to reflect both the similarity of node labels and the similarity of the local topology.
We first compute the score matrix $S_0$, based on the label similarity we described above, then we update our score to take into account the matching score of the parents of $n$ and $m$.

That way, $n$ will have a higher similarity score with $m$ if their respective parents are also similar.
We repeat the process $p$ times ($p$ for propagation) until we obtain a score function $S_{p}$ that reflects both the label similarity and the local topology similarity:
\begin{equation}\label{eq:score}
	S_{p}(n, m) = \sum_{i = 0}^{p} w_i S_0(parent^i(n), parent^i(m))
\end{equation}
where $parent^i(n)$ is the $i^{th}$ parent of $n$ (with $parent^0(n) = n$) and $w_0, w_1\dots w_{p}$ are weights.
In practice, to limit the complexity, we only compute $S_{p}$ for all couples that have a non-null initial score: $\{(n, m)\in T_1 \times T_2 | S_0 \neq 0\}$.

\subsubsection{Building the bipartite graph $G$ (step 3)}
Using our final score function $S_{p}$, we can now build the bipartite graph $G$: we iterate on all nodes $n \in T_1$ and we create an edge $e(n, m)$ for each node $m\in T_2$ where $S_{p}(n,m) \neq 0$ and associate it with the cost $c(n, m) = 1/(1+S_{p}(n, m))$.
Our resulting cost function is thus defined as follows:
\begin{equation}\label{eq:SFTM_cost}
c_{SFTM}(e) =
\begin{cases}
w_n,                    & \text{if }n_1\text{ or }n_2\text{ is a no-match node}\\
\frac{1}{1+S_{p}(n, m)},& \text{otherwise}
\end{cases}
\end{equation}
Importantly, unlike the bipartite graph built in the FTM algorithm, the resulting bipartite graph $G_{SFTM}$ is \emph{not complete} as only edges such that $S_{p}(n,m) \neq 0$ are considered. This is one of the key differences allowing to improve computation times.

\subsection{Implementation Details}\label{sec:implementation}
\begin{figure*}
\includegraphics[width=.8\linewidth]{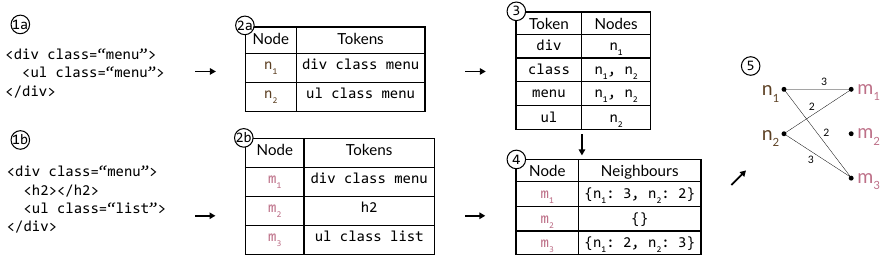}
\caption{Creating the bipartite graph $G$ from two example DOMs. (1a,b) are the input DOMs, (2a,b) the extracted tokens, (3) the inverted index $T_{map}$ , (4) the neighbours dictionaries and (5) the bipartite graph $G$. For simplicity, the figure shows a matching where IDF(t) = 1 , $p = 0$ and no-match nodes are not displayed.}\label{fig:inverted_index}
\end{figure*}

In the previous section, we introduced SFTM algorithm and described how it compares to FTM.
In this section, we describe more precisely how we implement the different steps of SFTM.  

\subsubsection{Node Similarity (step 1 and 2)}
Let us consider two trees $T_1$ and $T_2$.
We first build the dictionary $T_{map}$.
$T_{map}$ is an inverted index---\emph{i.e.}, each entry of $T_{map}$ is a tuple $(token, nodes)$ where $token$ is a token (usually a string) and $nodes$ is a \textit{set} of all $n \in T_1$ that belongs to $token$.
Figure~\ref{fig:inverted_index} (2a,b) depicts two examples of inverted index.
We note $T_{map}[key]$ the set of $nodes$ whose key in $T_{map}$ is $key$.
In Section~\ref{tokenSelection}, we further describe how we sort HTML nodes into tokens.

Given the inverted index $T_{map}$, we define the function $IDF: t \mapsto log(N/|T_{map}[t]|)$.
In order to limit the complexity of our algorithm, we remove every token $t \in T_{map}$ that is contained by more than $f(N) =\sqrt{N}$ nodes where $f$ is the chosen sublinear threshold function.
This is equivalent to putting a threshold on $IDF$ to only keep tokens $\{t \in T_{map} |IDF(t) > log(\sqrt{N})\}$.
Removing the most common tokens has a limited impact on matching quality since these are exactly the tokens that provide the least information on the nodes they appear in (see Figure~\ref{sec:parameter)}.

\begin{algorithm}
\SetAlgoLined
\KwIn{\\
m: a node in $T_2$\\
$T_{map}$: token map, dictionary of nodes from $T_1$ per token
}
\KwResult{neighbors: a dictionary of score per node in $T_1$}
$neighbors \gets new\ Dictionary()$\\
$tokens \gets tokens(m)$\\
\ForEach{t in tokens} {
    \ForEach{node in $T_{map}$[t]} {
        $neighbors[node] += IDF(t)$\\
    }
}
\Return neighbors
\caption{For a given node $m\in T_2$, compute similarity score $S_0(n, m)$ with all $n\in T_1$ such that $S_0 > 0$}\label{scoreAlgo}
\end{algorithm}

Once we have the token index $T_{map}$ and $IDF$, we apply Algorithm~\ref{scoreAlgo} on each node $m \in T_2$.
In Algorithm~\ref{scoreAlgo}, we first compute the tokens of the current node $m$ and for each token $t$, we use $T_{map}$ to retrieve the nodes $n \in T_1$ that contain the token $t$.
Each node $n$ thus retrieved is a \textit{neighbor} of $m$---\emph{i.e.}, $S_0(n, m) \neq 0$.
Finally, for each neighbour $n$ of $m$, we add $IDF(t)$ to the current score $S_0(n, m)$.
At this point, we have a $neighbors(m)$ dictionary for each node $m \in T_2$.
Each $neighbors(m)$ dictionary contains all non-null matching scores: $\forall n \in keys(neighbors(m)), neighbors(m)[n] = S_0(n, m)$.
Using the formula \ref{eq:score}, we can now easily compute $S_{p}$.

\subsubsection{Building the Token Vector}\label{tokenSelection}
The way we choose to compute the tokens contained in a node $n$ strongly influences the quality of our similarity score.
Finding the optimal way to compute these tokens has been the topic of numerous studies ~\cite{christen2011survey, steorts2014comparison, datar2004locality}.
We implemented the following $tokenize$ function to compute the tokens.
Given an HTML node $n$:
\begin{lstlisting}
<tag a[1]="v[1]" ... a[n]="v[n]">
 CONTENT
</tag> 
\end{lstlisting}
Where $p$ is the number of attributes, $(a[i], v[i]), i \in [1,l]$ are the attributes of $n$ and their associated values.
The absolute XPath of $n$ is $xPath(n)$, we say that $n$ contains the following tokens:
\begin{equation}
	tokenize(n) = \{tag, a[1]...a[l], tk(v[1])...tk(v[l]), xPath(n)\}
\end{equation}
where $tk$ is a standard string tokenizer function that takes a string and divides it into a list of tokens by splitting it on each non Latin character.

The absolute XPath of a node $n$ in a DOM is the full path from the root to the element where ranks of the nodes are indicated when necessary---\emph{e.g.}, \texttt{html/body/div[2]/p}.

\subsubsection{Building G (step 3)}
Using Equation~\ref{eq:SFTM_cost}, we compute the cost $c(n, m)$ for each couple $(n,m)$ where $S_p(n,m) \neq 0$.
Then, for each node $m\in T_2$, we add one edge for all nodes $neighbours(m) \in T_1$ .

\subsubsection{Metropolis Algorithm (step 4)}
Once we built the graph $G$ with its associated costs, we need to find the set of edges $M$ in $G$ that constitutes the best full matching.
In order to do so, we use the same technique as FTM.
But, when it comes to applying the Metropolis algorithm, SFTM differentiates from FTM in two ways: 
\begin{inparaenum}
	\item we modified the objective function and
    \item SFTM matching suggestion function is faster to compute since costs never need to be recomputed.
\end{inparaenum}

FTM uses the objective function $f_{FTM}(M) = \exp(-\beta\ c(M))$.
In the original FTM paper, the authors noted that the parameter $\beta$ seemed to depend on $|M|$.
In order to avoid this dependency, we normalize the total cost:
\begin{equation}
	f_{SFTM}(M) = \exp(-\beta\ \frac{c(M)}{|M|})
\end{equation}
The function $suggestMatching: M_t \mapsto M_{t+1}$ takes a full matching $M_t$ and returns a full matching $M_{t+1}$ related to $M_t$.
In the following Algorithm~\ref{suggestMatching}, 
\begin{compactenum}
	\item $selectEdgeFrom(edges)$ loops through $edges$ (in order) and at each iteration $i$, has a chance $\gamma \in [0,1]$ to stop and return $edges[i]$,
    \item $connectedEdges(edge)$, where $edge$ connects $u$ and $v$, returns the set $E$ of all edges connected to $u$ or $v$ (note that $edge \in E$).
\end{compactenum}

\begin{algorithm}
\SetAlgoLined
\KwData{$G$ : The bipartite graph}
\KwIn{$M_t$: A full matching}
\KwResult{$M_{t+1}$: the suggested full matching}
 $M_{t+1} \gets []$  \\
 $remainingEdges \gets sortedEdges(G)$ \\
 $toKeep \gets randomInt(0, |M_t|)$ \\
 \For{i = 0 .. toKeep}{
 	$M_{t+1}$.add(edge)\\
    remainingEdges.removeAll(connectedEdges(edge))
 }
 \While{$remainingEdges$ is not empty} {
    $edge \gets SelectEdgeFrom(remainingEdges)$ \\
    $M_{t+1}$.add(edge)\\
    remainingEdges.removeAll(connectedEdges(edge))
 }
 \Return $M_{t+1}$
 \caption{Suggest a new matching}\label{suggestMatching}
\end{algorithm}

In practice, we first compute all the connected nodes and edges before storing them as dictionaries, so that the function $connectedEdges$ in Algorithm~\ref{suggestMatching} can be computed in $O(1)$ time.
It is worth noting that, to allow fast removal, the list $remainingEdges$ is implemented as a double-linked list.
The parameter $\gamma$ defines a trade-off between exploration (low $\gamma$) and exploitation (high $\gamma$).

\subsection{Complexity Analysis}\label{sec:complexity}
We are interested in evaluating the time complexity of the algorithm with respect to $N$.
In our analysis, we consider that $n_t$, the maximum number of tokens per node is a constant since it does not evolve with $N$.

When building $G$, we first compute the inverted index $T_{map}$.
Computing $T_{map}$ requires to iterate through tokens of all nodes in $T_1$, which implies a complexity in $O(N \cdot n_t) = O(N)$.

To find the neighbours of nodes from $T_2$ using $T_{map}$, we iterate through all the nodes in $T_2$.
Each node in $T_2$ has $n_t$ tokens.
The number of nodes containing a token is artificially limited to $f(N)$.
Thus, building the similarity function $S_0$ takes $O(N \cdot f(N))$ time.

For each $m\in T_2$, we create an edge for each neighbor $n \in T_1$.
Each token $t \in m$ adds up to $f(N)$ neighbors.
It means that the total number of edges is in $O(N \cdot n_t \cdot f(N))$ = $O(N \cdot f(N))$. 

Before executing the Metropolis algorithm on $G$, we sort the edges by cost, which takes $O(N \cdot f(N) \cdot log(N \cdot f(N))) = O(N \cdot f(N) \cdot log(N))$ (as $f(N) \leq N$).
Finally, at each step of the Metropolis algorithm, we run the $suggestMatching$ function, which prunes a maximum of $O(f(N))$ neighbors for each one of the $N$ edges it selects.

Overall, sorting all edges requires the highest theoretical complexity: $O(N \cdot f(N) \cdot log(N))$.
If no threshold is set---\emph{i.e.} $f(N) = N$---then the overall complexity of SFTM is $O(N^2 \cdot log(N))$, which keeps outperforming the TED ($O(N^3)$) and FTM ($O(N^4)$). 

In the evaluation, we used $f(N) = \sqrt{N}$ which leads to a complexity in $O(N \cdot \sqrt{N} \cdot log(N))$.
The empirical evaluation conducted in Section~\ref{sec:evaluation} tend to suggest that our analysis might be too pessimistic.

\section{Empirical Evaluation}\label{sec:evaluation}
The objective of this evaluation is to assess that:
\begin{compactenum}
	\item The quality of the matchings computed by SFTM compares with the baseline APTED,
    \item The SFTM algorithm offers practical speed gains on real-life web documents.
\end{compactenum}

\subsection{Input Web Document Dataset}
We need to assess the ability of SFTM to match the nodes between two slightly different DOM $D$ and $D'$.

\paragraph{DOM mutation.}
To build a dataset of $(D,D')$ tuples where the ground truth (perfect matching) is known, we developed a mutation-based tool that works the following way:
\begin{compactenum}
	\item We construct the DOM $D$ from an input web document,
    \item For each element of $D$, we generate a unique signature attribute,
    \item For each original DOM $D$, we randomly generate a set of mutated versions: the \textit{mutants}.
    Each \textit{mutant} $D'$ is stored along with the precisely described set of mutations that was applied to $D$ to obtain $D'$.
    Importantly, the signature tags of the elements in $D$ are transferred to $D'$, which constitutes the perfect matching between $D$ and $D'$.
\end{compactenum}

In our tool, most attention has been dedicated to the choice of relevant mutations to apply.
The following table summarizes the set of relevant mutations possibly applied to an element of the DOM.
\begin{center}
   \begin{tabular}{|l l|} 
       \hline
       Element       & Mutation operators\\
       \hline
       \hline
       \em Structure & \tt remove, duplicate, wrap, unwrap, swap \\
       \em Attribute & \tt remove, remove words \\
       \em Content   & \tt replace with random text, change letters, \\
                     & \tt remove, remove words \\
       \hline
  \end{tabular}
\end{center}

\paragraph{Baseline algorithms.}
We compare SFTM to APTED, which is the reference implementation of TED that yields the best performance so far.\footnote{We also implemented the original FTM, but the computation times and space complexity of this implementation were too high to run the algorithm on real-life web documents (\emph{e.g.} on a toy example with 58 nodes, the computation took 1 hour).}
The implementation of APTED used for the evaluation is the one provided by the authors of~\cite{pawlik2016tree,pawlik2015efficient}.
We consider the pairs $(D, D')$ taken from the above web document dataset, and we ran SFTM and APTED algorithms with each pair to match $D$ with $D'$ on the same machine.

\paragraph{Input document sample.}
We fed our mutation tool with the home pages of the Top\,1K Alexa websites.
For each DOM $D$ thus retrieved, we created 10 mutants $D'$ with a number of mutated nodes ranging from $0$ to $50\%$ of the total number of nodes on the page.

\begin{figure}
    \centering
    \includegraphics[width=\linewidth]{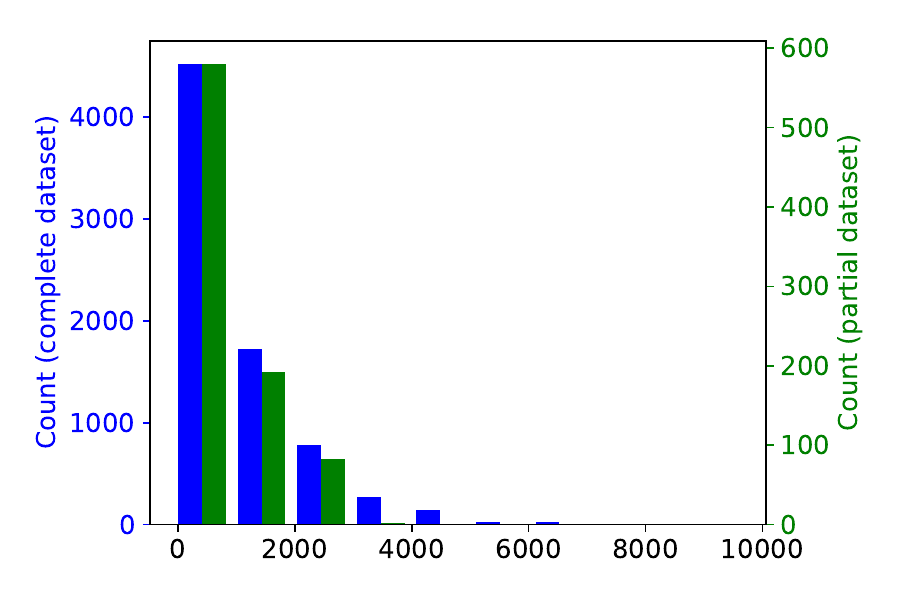}
    \caption{The distribution of DOM sizes (in terms of nodes) for the complete (left, blue) and the partial (right, green) datasets.}
    \label{fig:distribution}
\end{figure}

Overall, we considered an input dataset composed of $7,502$ document tuples.
We ran SFTM on the complete dataset but, due to high computation times, APTED can only be evaluated on a subset of this dataset comprising $852$ tuple documents, which represents a 3\,\% error margin with 95\,\% confidence with respect to the complete dataset.
Figures~\ref{fig:quality} and~\ref{fig:rted_vs_sftm} comparing APTED and SFTM are based on this partial dataset, while the complete dataset was used when studying SFTM in isolation (cf. Figure~\ref{fig:sftm_speed}).
Figure~\ref{fig:distribution} reports on the size distribution, in number of nodes, of the selected web documents for both complete and partial datasets.

\paragraph{Ground truth.}
When building the dataset, we keep track of nodes' signature so that we always know which nodes from $D$ should match with nodes from $D'$.
This ground truth is ignored by the evaluated algorithms, but is used \emph{a~posteriori} to measure and compare the quality of the matchings computed by the algorithms under evaluation.

\subsection{Experimental Results}\label{sec:performances}
\begin{figure}
  \centering
  \includegraphics[width=\linewidth]{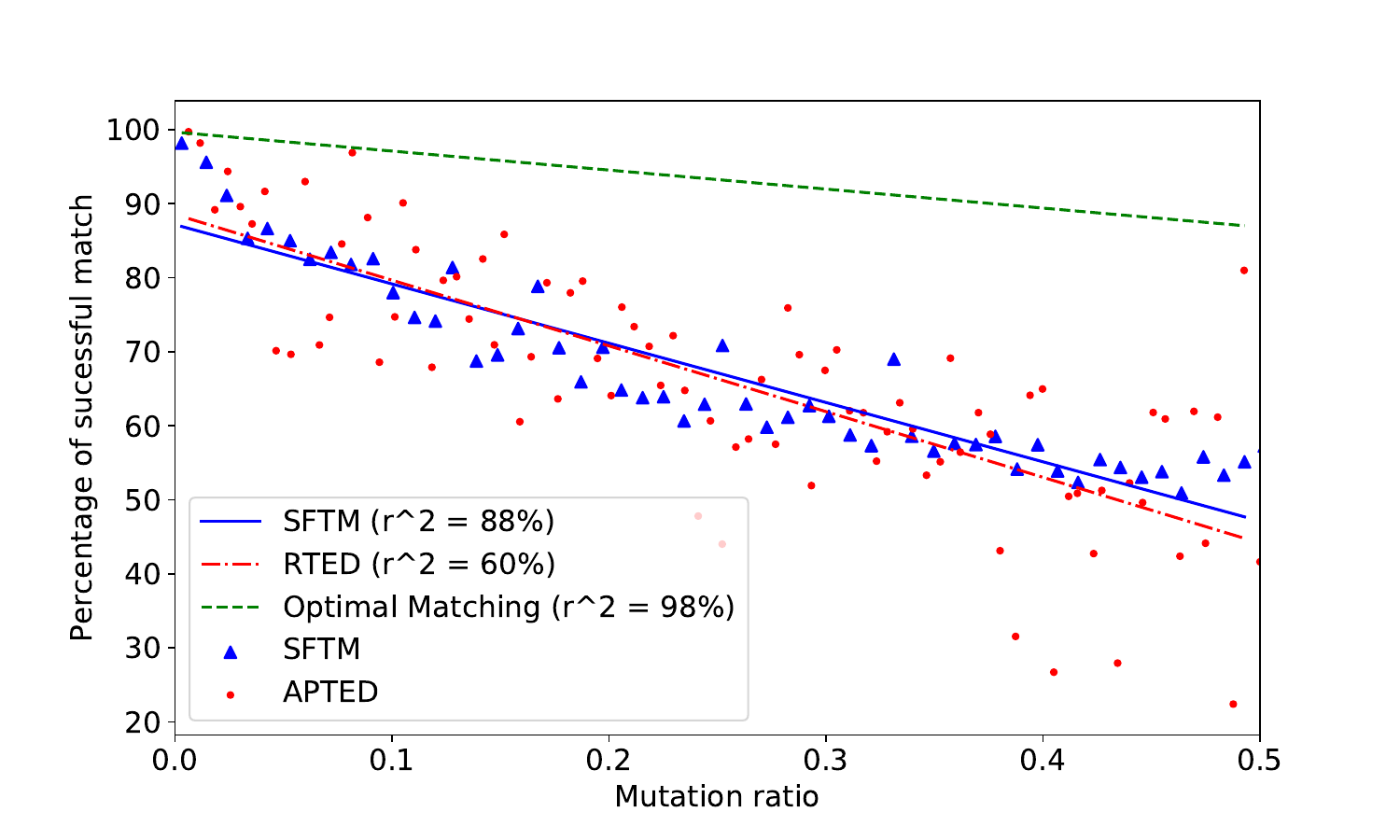}
  \caption{Successful match rate according to mutation ratio for SFTM and APTED}
  \label{fig:quality}
\end{figure}

\paragraph{Matching quality.}
The signature tags on nodes from $D$ and $D'$ allow us to judge the quality of the matching according to two metrics: 
\begin{inparaenum}
	\item \textit{mismatch}, the number of nodes couples that were wrongly matched and
    \item \textit{no-match}, the number of nodes from $D$ that were matched with no nodes from $D'$.
\end{inparaenum}
We call \textit{successful match rate}, the number of couples rightfully matched by the algorithm---\emph{i.e.}, that is neither a \textit{mismatch} nor a \textit{no-match}. 
The list of possible mutations between $D$ and $D'$ include the removal of a node.
In case we remove a node, the algorithms will (legitimately) not be able to match the removed nodes.
We call \textit{optimal successful match rate}, the maximum ratio of nodes that the algorithms can successfully match: $\frac{total\ nodes\ in\ D - number\ of\ removals\ in\ D'}{total\ nodes\ in\ D}$.
To measure the quality of the matchings, we compare the successful match rate of the matchings computed by both algorithms with the optimal successful match rate on Figure~\ref{fig:quality}.

We observe that SFTM and APTED have very similar performance.
They both seem to perform linearly with the number of mutations.
However, APTED is much less stable than SFTM with a correlation coefficient $r^2 = 60\%$.

\begin{figure}%
        \centering
        \includegraphics[width=\linewidth]{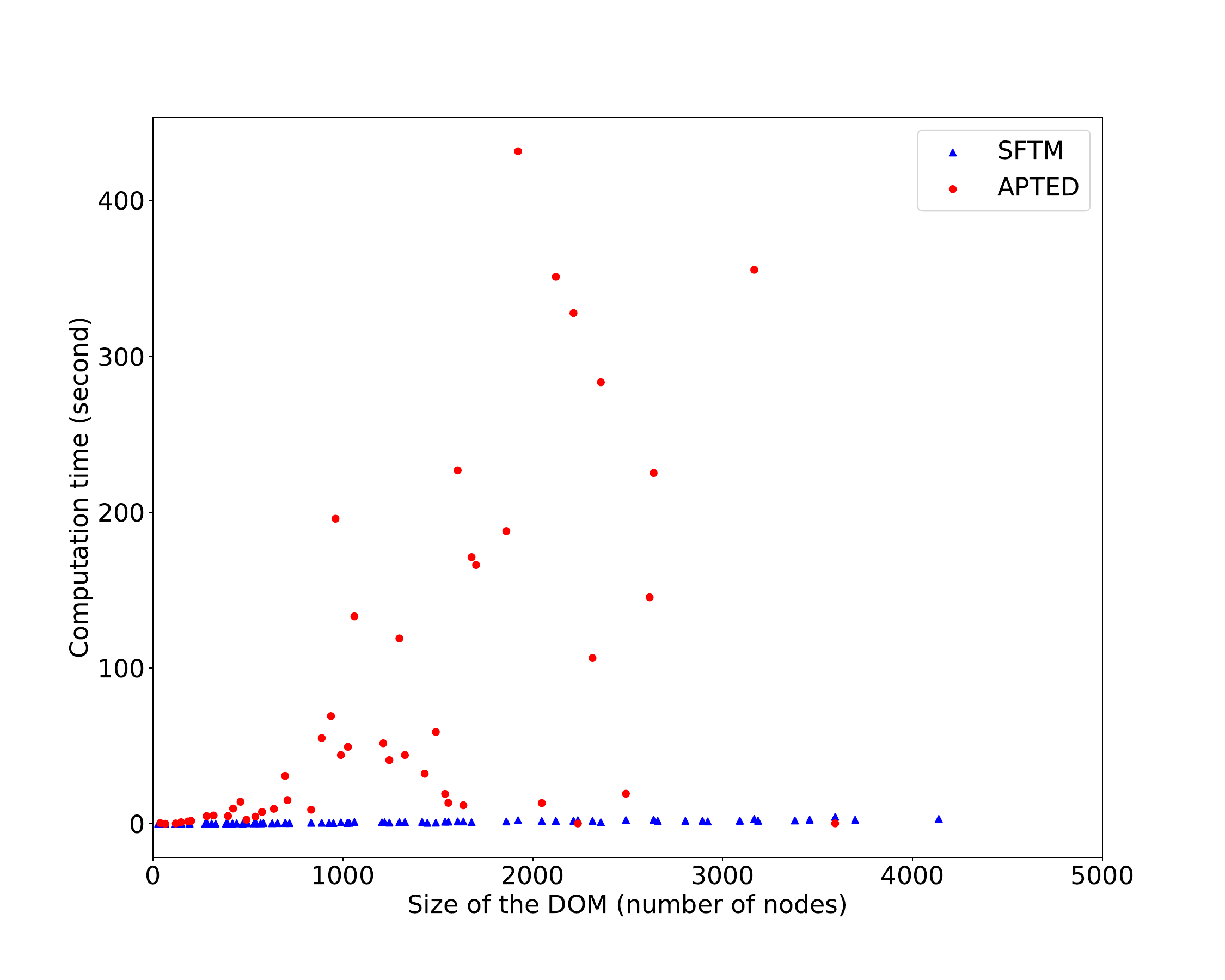}
        \caption{Average time to compute matching by DOM size for APTED and SFTM. APTED computation timeout has been set at 450 seconds}
        \label{fig:rted_vs_sftm}
    \end{figure}

\paragraph{Completion time.}
For each couple ($D$, $D'$) retrieved from the dataset, we measured the time taken by SFTM and APTED to compute a matching.
For practical reasons, we set a timeout to APTED computations at 7 minutes (450 seconds).
Figure~\ref{fig:rted_vs_sftm} reports on the average time (in seconds) to match DOM couples of increasing size (in terms of number of nodes) for both algorithms.
We note that APTED computation time varies greatly depending on the DOM couple.
While the theoretical worst case complexity of APTED is $O(n^3)$, we can observe in practice that APTED may run up to 100 times slower than SFTM.

\begin{figure}%
        \centering
        \includegraphics[width=\linewidth]{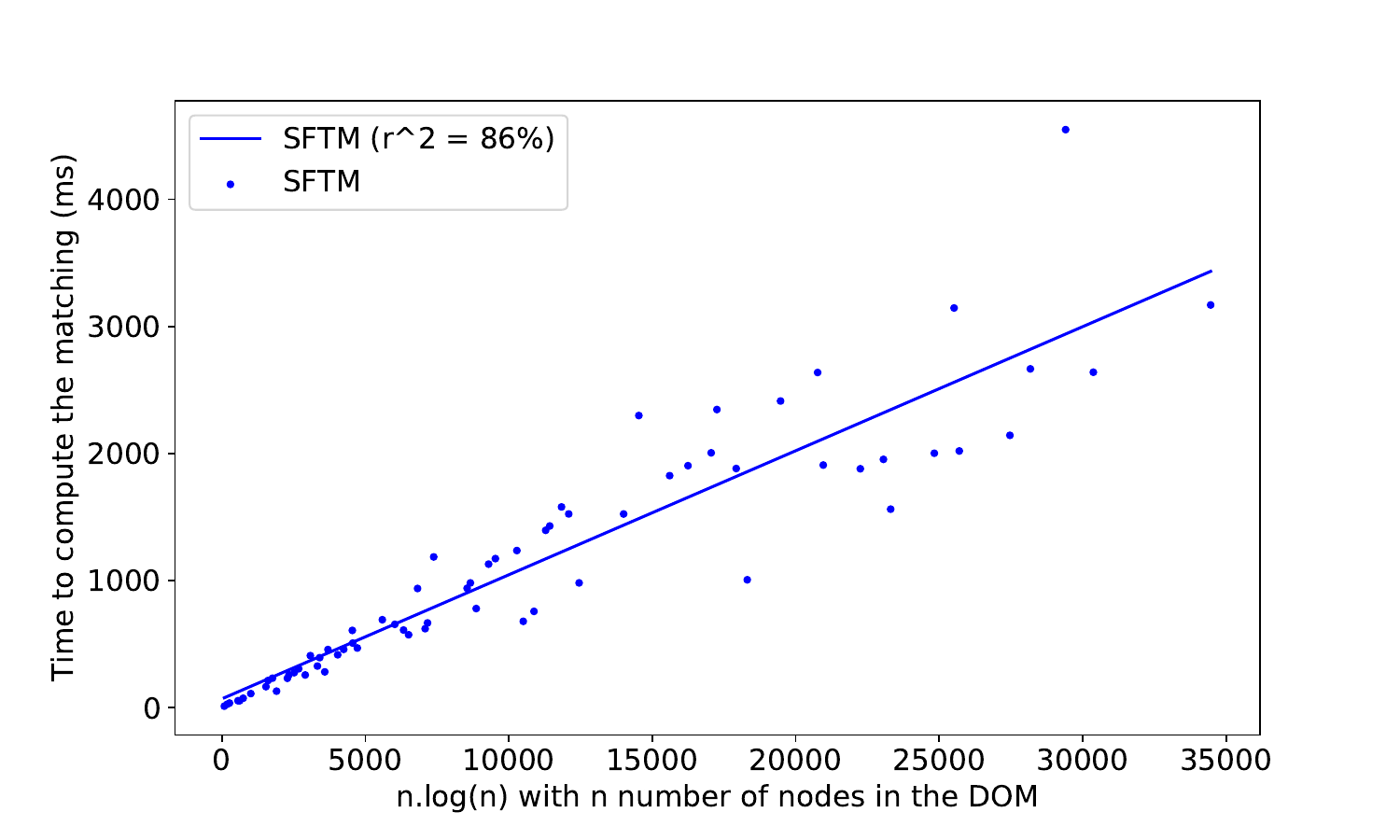}
        \caption{Average time to compute matching by  $n\ log(n)$ value for SFTM, with $n$ the number of nodes in the DOM}
        \label{fig:sftm_speed}
    \end{figure}

Figure~\ref{fig:sftm_speed} delivers a closer look on the scalability of SFTM.
The empirical results seem to indicate an evolution in $O(n \cdot log(n))$: in  Figure~\ref{fig:sftm_speed}, we replaced the X axis from the number of nodes in the DOM $n$ to $n\ log(n)$ then computed a linear regression on the curve which resulted in a correlation coefficient $r^2 = 86\%$.

This observation raises the question of the impact of the sublinear threshold function $f$ on the performance of SFTM.
We therefore conducted a sensitivity analysis of this parameter to better understand potential trade-offs offered by the definition of this function, with regards to the complexity analysis we performed (cf. Section~\ref{sec:complexity}).

\paragraph{Parameter sensitivity.}
Since we aim at improving the performances of SFTM in term of computation times, we study the sensitivity of the sublinear threshold function $f$ which is a parameter that directly influences the computation time of the algorithm.

\begin{figure}
    \centering
    \includegraphics[width=\linewidth]{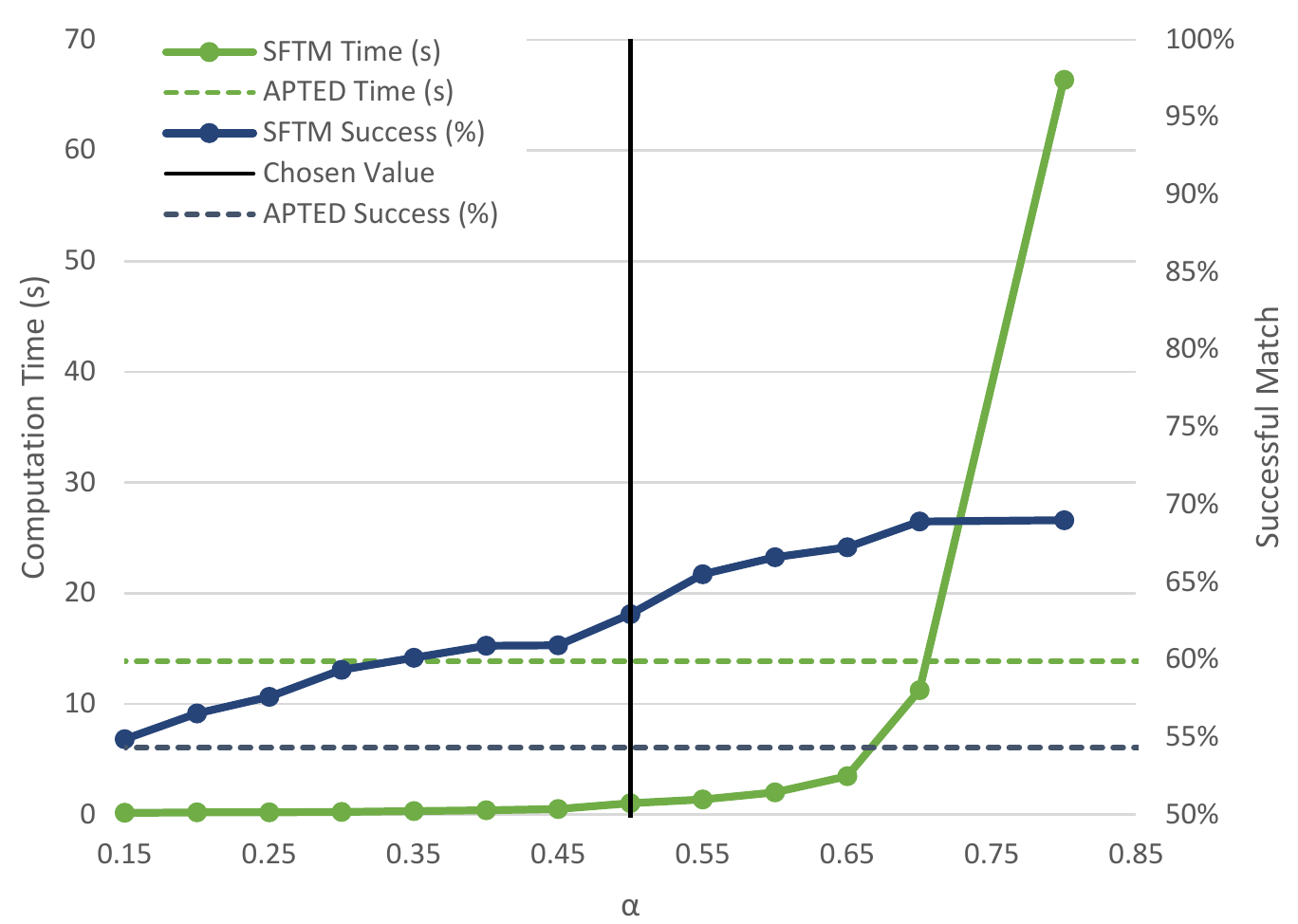}
    \caption{Performance of SFTM given $f(N) = N^\alpha$ according to $\alpha$. APTED performance on the same dataset is shown as a reference}
    \label{fig:parameter}
\end{figure}

Figure~\ref{fig:parameter} reports on the evolution of SFTM performances when $f$ varies.
To study the sensitivity of $f$, we choose to use the power function $f(N) = N^\alpha$ as a threshold and display how the computation times and matching accuracy evolve with $\alpha$.

For this experiment, as we are interested in studying the sensitivity of the $\alpha$ parameter on the performances of SFTM, we consider a subset of $53$ tuples from the complete dataset used in previous sections (cf. Section~\ref{sec:performances}), which represents a 13.4\,\% error margin with 95\,\% confidence.
On average, on this subset, DOM trees contain $1,065$ nodes and mutants have a $22\%$ mutation ratio, which remains representative of the complexity of web documents considered in this paper.

As expected, when $\alpha$ increases, the quality of the matching and the computation times increase.
However, beyond a certain value of $\alpha$, the increase of computation time is significantly superior to the gain in accuracy: increasing $\alpha$ from $0.5$ to $0.8$ entails more than 60 times longer computation times for only $3.5\%$ gain in accuracy. 
Intuitively, this is because tokens contained in most nodes provide very few information (low \emph{Inverse Document Frequency}), but increase the complexity quadratically.
In this paper, we used $\alpha = 0.5$ (\emph{i.e.}, $f(N) = \sqrt{N}$): this value achieves good enough performances to demonstrate that SFTM can match two real-life web documents in practical time without compromising on quality.  

\section{Threats to Validity}\label{sec:threats}
The absolute values of completion times depend on the machine on which the algorithms were executed.
As computations took time, we had to run both SFTM and APTED on a server, which is shared among several users.
Although we paid a careful attention to isolate our benchmarks, the available resources of the server might have varied along execution thus impacting our results.
Nevertheless, the repetition of measures reports a clear signal in favour of SFTM.

Our dataset contains the homepages of the Top\,1k Alexa websites.
The fact that our qualitative evaluation has only been conducted on homepages might have biased the results since such pages might not be fully representative of the complexity of online documents.

We evaluated the quality of the matchings using synthetic mutations on real-life websites.
We dedicated a lot of thought into choosing an objective set of potential mutations representative of real-life evolution of websites.
However, there is still a chance we missed some common mutations to which SFTM might prove to be not robust.

\section{Conclusion \& Perspectives}\label{sec:conclusion}
Comparing modern real-life web documents is a challenge for which traditional \emph{Tree Edit Distance} (TED) solutions are too restricted and computationally expensive. 
\cite{Kumar2011_FTM} introduced \emph{Flexible Tree Matching} (FTM) to offer a restriction-free matching, but at the cost of prohibitive computational times.
In this paper, we presented \emph{Similarity-based Flexible Tree Matching} (SFTM), which extends FTM to offer tractable computational times while offering non-restricted matching.
We evaluated our solution using mutations on real-life documents and we showed that SFTM qualitatively compares to TED while improving the performances by two orders of magnitude.
The proof of concept we deliver demonstrates that matching real-life web documents in practical time is possible.

We believe that having a robust algorithm to efficiently compare web documents will open up new perspectives within the web community.
In future work, we will further investigate on how to improve the quality of the matchings by analyzing which situations cause SFTM to make mistakes in order to establish guidelines to adjust the exposed parameters.

Whether our work might be applicable to other trees than web DOMs remains to be tested.
Indeed, SFTM strongly relies on the fact that node labels in DOMs are highly differentiating (many specific attributes on each element), which is not the case for all kinds of trees.

\bibliographystyle{ACM-Reference-Format}
\bibliography{references}

\end{document}